\documentclass[%
 reprint,superscriptaddress,
amsmath,amsfonts,amssymb,
 aps,prl,showkeys,
]{revtex4-1}
\usepackage{endnotes}
\usepackage{graphicx}
\usepackage{mathtools}
\usepackage[makeroom]{cancel}
\usepackage{amssymb}
\usepackage{dcolumn}
\usepackage{fullpage}
\usepackage{amsmath}
\usepackage{multirow}
\usepackage{physics}
\usepackage{dcolumn}
\usepackage{bm}

\begin{document}

\title{Tunable Circular Dichroism and Valley Polarization\\
 in the Modified Haldane Model}

\author{Marc Vila}
\affiliation{Catalan Institute of Nanoscience and Nanotechnology (ICN2), CSIC and BIST,
Campus UAB, Bellaterra, 08193 Barcelona, Spain}
\affiliation{Department of Physics, Universitat Aut\`onoma de Barcelona, Campus UAB,
Bellaterra, 08193 Barcelona, Spain}
\author{Nguyen Tuan Hung }
\affiliation{Department of Physics, Tohoku University, Sendai 980-8578, Japan}
\author{Stephan Roche}
\affiliation{Catalan Institute of Nanoscience and Nanotechnology (ICN2), CSIC and BIST,
Campus UAB, Bellaterra, 08193 Barcelona, Spain}
\affiliation{ICREA--Instituci\'o Catalana de Recerca i Estudis Avan\c{c}ats, 08010 Barcelona, Spain}
\author{Riichiro Saito}
\affiliation{Department of Physics, Tohoku University, Sendai 980-8578, Japan}

\begin{abstract} 
We study the polarization dependence of optical absorption for the modified Haldane model, which exhibits antichiral edge modes in presence of sample boundaries and has been argued to be realizable in transition metal dichalcogenides or Weyl semimetals. A rich optical phase diagram is unveiled, in which the correlations between perfect circular dichroism, pseudospin and valley polarization can be tuned independently upon varying the Fermi energy. Importantly, perfect circular dichroism and valley polarization are achieved simultaneously, a feature not yet observed in known optical materials. This unprecedented combination of optical properties suggests some interesting novel photonic device functionality (e.g. light polarizer) which could be combined with valleytronics applications (e.g. generation of valley currents).
\end{abstract}

\maketitle

Circular dichroism (CD) is the ability of a material to exhibit a significant differential absorption of left- and right-handed circularly polarized light (LCP and RCP). This optical property appears in chiral materials and molecules, and presents large interest for biopharmaceutical applications, as well as molecular spectroscopy or even quantum information processing \cite{Oh2015,Wang2016,He2016,Huang18}. However, CD is generally too small for practical applications \cite{Wang2016review, Sato2017} so that there is a great interest in exploring if novel materials such as carbon nanotubes \cite{Yokoyama2014} or metamaterials \cite{Wang2016, Wang2016review} cound manifest stronger CD. On the other hand, two-dimensional materials, ranging from insulating hexagonal boron nitride (hBN) and semiconducting transition metal dichalcogenides (TMDs) to semimetallic graphene, display optical properties that differ from their bulk parental materials \cite{Xiao2012,Xia2014}. TMDs exhibit particularly strong light absorption, and owing to the broken inversion symmetry and time reversal invariance, light with opposite handedness leads to preferential population of the K or K' valleys of the Brillouin zone. Such effect is known as valley polarization (VP) \cite{Ghalamkari2018b} or valley-selective circular dichroism \citep{Cao2012}, and it is of paramount importance to realize valleytronic applications \cite{Schaibley2016,Vitale2018}. Strong VP and large quantum efficiency have been achieved with TMDs \cite{Cao2012,Mak2012,Zeng2012,Withers2015,Xinlin2016}. Noteworthy, this valley polarization is not restricted to TMDs, but to any hexagonal lattice with broken inversion symmetry \cite{Ghalamkari2018b}, including hBN or novel graphene-based materials \cite{Liu2018}. Nevertheless, the presence of VP as an intrinsic property suggests that CD is not simultaneously achievable, since both LCP and RCP light are equally absorbed. Merging these two properties together would result in novel devices possessing both photonic and valleytronic functionalities. Several attempts on inducing CD in TMDs have been reported \cite{Purcell2018,Lin2018}, but the effect was produced by external chiral molecules. Therefore, a two-dimensional system inherently presenting both valley polarization and circular dichroism is still missing.   

A paradigm for a 2D lattice model was introduced by Haldane \cite{Haldane1988}. Together with the Kane and Mele model \cite{Kane05}, these models have pioneered the development of time-reversal invariant topological insulators  \cite{Hasan2010}. Very recently, Colom{\'e}s and Franz have proposed a {\it modified Haldane model} exhibiting dispersive “antichiral” edge states, which are modes propagating in the same direction at both parallel ribbon edges and compensated by bulk counterpropagating modes \cite{Colomes2018}. Such edge states, resilient to disorder, are believed to resemble the Fermi arcs in 3D Dirac and Weyl semimetals and could be useful for low-power electronics or spintronics. While the experimental realization of the Haldane model has been achieved using ultracold atoms in an optical lattice \cite{Jotsu2010}, a realization of the modified Haldane version might be possible with Weyl semimetals, TMD monolayers \cite{Xiao2012, Zhu2011, Armitage2018, Colomes2018, Tong2016} or ferromagnet/graphene/TMD heterostructures \cite{Frank2018, mHaldane}. Besides, given that the Haldane model has recently been predicted to exhibit perfect CD when the system is topological nontrivial \cite{Ghalamkari2018}, the exploration of the fundamental optical properties of the modified Haldane model might help to realize circular dichroism in presence of valley polarization. 

In this Letter, we show that the phase diagram of the modified Haldane model presents the possibility for simultaneous realization of valley polarization together with a \textit{perfect} circular dichroism.  More remarkably, the combination of perfect VP and CD can be monitored by varying the Fermi energy, while the tuning of the modified Haldane parameters ($\{ \Delta,t_2,\phi\}$, see Eq. (\ref{Hm})) allows the description of a broad class of materials such as semimetals and indirect semiconductors. The modified Haldane model is defined on a honeycomb lattice with broken inversion and time reversal symmetries. Its tight-binding version reads \citep{Colomes2018}:
\begin{equation}
\mathcal{H}=t_1 \sum_{\langle i,j \rangle} c_i^\dagger c_j + t_2 \sum_{\langle \langle i,j \rangle \rangle} e^{-i \nu_{ij}\phi} c_i^\dagger c_j + \Delta \sum_i \mu  c_i^\dagger c_i.
\label{Hm}
\end{equation}

\begin{figure}[t]
\centering
\includegraphics[width=1 \columnwidth]{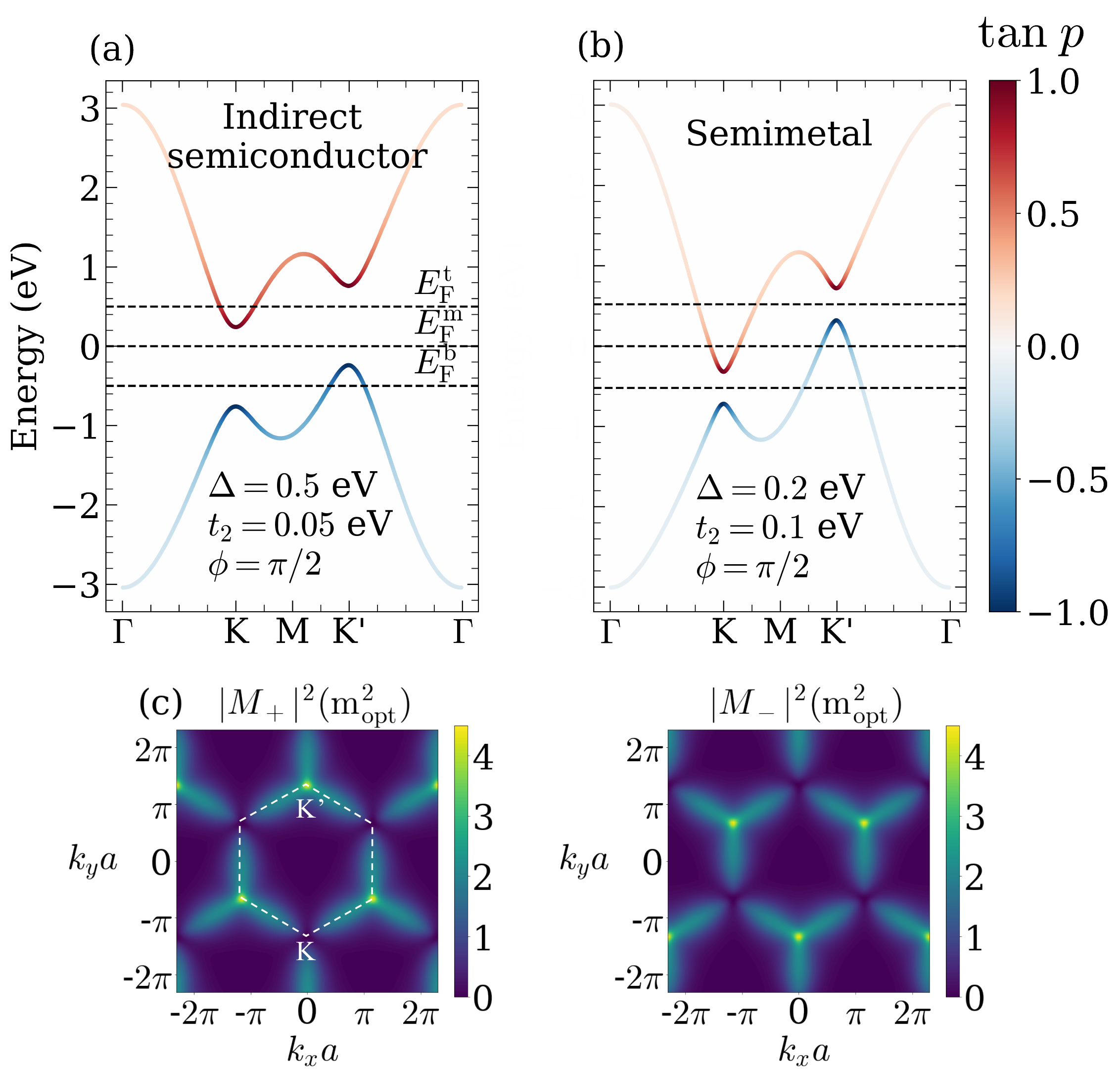}
\caption{Band structure of the modified Haldane model for a given set of parameters that produces an (a) indirect semiconductor and a (b) semimetallic phase. In all cases, $t_1 = 1$ eV. The value of the pseudospin polarization $\tan p$ (Eq. (\ref{tanp})) is given in color scale. The horizontal dashed lines highlight selected Fermi energies at which $\tan V_\sigma$ (Eq. (\ref{tanV})) and $\tan \theta$ (Eq. (\ref{CD})) are calculated in Figs. (2) and (\ref{fig:fig3}). (c) $|M_\sigma(\mathbf{k})|^2$ in units of the atomic dipole vector, for the parameters used in (b), over the full Brillouin zone (superimposed white, dashed line) for LCP light (left) and RCP light (right).}
\label{fig:fig1}
\end{figure}

The first term is the nearest-neighbor hopping with magnitude $t_1$ and $c_i^\dagger$ ($c_{j}$) is the creation (annihilation) operator for a spinless electron on site $R_{i}$($R_{j}$) of the honeycomb lattice. The second term is a next-nearest-neighbor (NNN) hopping with complex value $t_2e^{-i\nu_{ij}\phi}$ that breaks time reversal symmetry \cite{Haldane1988, Colomes2018}. In the Haldane model $\nu_{ij} = \pm1$  with +1 for counterclockwise (ccw) and -1 for clockwise (cw) hoppings. In contrast, in the modified Haldane model $\nu_{ij}=\pm 1$ for sublattice A while $\nu_{ij}=\mp 1$ for sublattice B, which besides breaking time reversal symmetry, also introduces a scalar potential with an opposite sign in each valley, which is crucial for obtaining both antichiral edge states \cite{Colomes2018} and optical absorption tunability, as discussed herebelow. The third term breaks inversion symmetry when the potential is different for the sublattices A ($\mu=+1$) and B ($\mu=-1$) and opens a gap with value $E_\text{{gap}}=2\Delta$ at the K/K$^\prime$ points. It is also useful to express Eq. (\ref{Hm}) in reciprocal space and expand it near the vicinity of K/K$^\prime$ up to first order in momentum $\mathbf{q}$:
\begin{align}\label{mH_K}
\mathcal{H}=\hbar v_{F}(\kappa \sigma_{x}q_{x}+ \sigma_{y}q_{y})-t_{2}^{a} \sigma_{0} - \kappa  t_2^b  \sigma_{0}+\Delta  \sigma_{z},
\end{align}
with $v_F$ being the Fermi velocity, $t_{2}^{a}=3t_2\cos\phi$, $t_2^b=3\sqrt{3}t_2\sin\phi$ and $\kappa=+1 (-1)$ for the K (K$^\prime$) valley. Here, $\sigma_0$ is the $2 \times 2$ identity matrix and $\sigma_i$ $(i = x, y, z)$ are the Pauli matrices acting on the sublattice space. The energy dispersion results in $\varepsilon(\mathbf{q}) = -t_{2}^{a} - \kappa t_{2}^{b} \pm \sqrt{\Delta^2+(\hbar v_F \mathbf{q})^2}$, where the $\pm$ sign stands for the conduction band/valence band. As mentioned previously, the NNN hoppings results in a potential opposite in each valley, which depends on the terms $t_2$ and $\sin\phi$ (third term in Eq. (\ref{mH_K})). However, although the valleys are relatively shifted, the gap is the same and only depends on $\Delta$. We plot the band structure in Fig. \ref{fig:fig1}(a) and Fig. \ref{fig:fig1}(b) for two different set of parameters $\Delta$, $t_2$ and $\phi$. In both cases, we identify the Dirac cone shift and gap opening, but whereas Fig. \ref{fig:fig1}(a) shows the case of an indirect semiconductor, Fig. \ref{fig:fig1}(b) reveals a semimetallic phase. This indirect semiconductor to semimetallic transition occurs when $|\frac{\Delta}{t_2}| \leq 3\sqrt{3} \, |\sin\phi|$. 

The optical properties of the modified Haldane model are calculated based on the perturbed electron-photon Hamiltonian in the dipole approximation \cite{Gruneis2003,Ghalamkari2018,Ghalamkari2018b} and using the Fourier transform of Eq. (\ref{Hm}). The optical absorption transition probability from an initial state in the valence band $v$ to a final state in the conduction band $c$ depends on the squared optical matrix elements $|M_\sigma(\mathbf{k})|^2= | \mathbf{P}\cdot \mathbf{D}(\mathbf{k}) |^2$. Here, $\mathbf{P}$ is the polarization vector ($\textbf{P}=\frac{1}{\sqrt{2}}(1\ i\sigma)^T$) and $\mathbf{D}(\mathbf{k})$ stands for the dipole vector, that has the form $\langle c | \nabla | v \rangle$. We consider only the case of circularly polarized light propagating in the direction perpendicular to the lattice, whose polarization is defined by $\sigma=+1 (-1)$ for LCP (RCP) light. 

Using the parameters of Fig. \ref{fig:fig1}(b), we numerically calculate the optical matrix elements in the whole Brillouin zone for LCP and RCP. The result is plotted in Fig. \ref{fig:fig1}(c). One sees that $|M_\sigma(\mathbf{k})|^2$ presents valley polarization as in other 2D hexagonal materials with broken inversion symmetry \cite{Ghalamkari2018,Xiao2012}. While LCP (RCP) is maximum at the K$^\prime$ (K) valley, it is 0 at the K (K$^\prime$). The VP was related to different signs of the Berry curvature at different valleys due to the $\Delta$ term \cite{Yao2008,Xu2014}. We reach the same conclusion by noting that $\Delta$ produces pseudospin polarization (PP), defined as the difference of wavefunction localization between atom A and atom B. This PP is encoded in the wavefunction and consequently also in $\mathbf{D}(\mathbf{k})$ and $|M_\sigma(\mathbf{k})|^2$, resulting in VP. Thus, it is insightful to calculate the pseudospin polarization as it is related to the optical absorption at the K and K$^\prime$ valleys. We define PP in analogy to the circular dichroism (see Eq. (\ref{CD}) and Ref. \cite{Sato2017,Ghalamkari2018}) as

\begin{equation}\label{tanp}
\tan p=\frac{|C_A^d(\textbf{k})|^2-|C_B^d(\textbf{k})|^2}{|C_A^d(\textbf{k})|^2+|C_B^d(\textbf{k})|^2}, \quad (d = v, c).
\end{equation}

Here, $C_A^d(\textbf{k})$ and $C_B^d(\textbf{k})$ are the coefficients of the Bloch wavefunction in band $d$ of sublattice A and B, respectively, which we obtain by numerical diagonalization of the Hamiltonian. The PP is color-plotted in the bands of Fig. \ref{fig:fig1}(a) and Fig. \ref{fig:fig1}(b). We observe that the PP is opposite for the valence and conduction bands and equal for both valleys, resulting in VP \cite{Ghalamkari2018}. Moreover, the PP is maximum at K and K$^\prime$ which explains why the absorption is also maximum there as seen in Fig. \ref{fig:fig1}(c). Using the notation ($c$, $v$) we obtain $\tan p = (+1, -1)$ for positive $\Delta$, but when $\Delta$ is negative, $\tan p = (-1, +1)$ and the VP reverses (LCP (RCP) is absorbed at K (K$^\prime$)). Finally, for $\Delta = 0$, there is no pseudospin polarization and both valleys absorb both LCP and RCP lights. We note that the wavefunction does not depend on $t_2$ or $\phi$ and therefore $|M_\sigma(\mathbf{k})|^2$ is not affected by the scalar potential that shifts the Dirac cones.

Next, we show that by changing the Fermi energy, the modified Haldane model also exhibits perfect CD concurrently with VP. To that end, we now compute the optical absorption with the Fermi's golden rule considering the occupation probability of states in the valence and conduction bands with the Fermi-Dirac distribution ($f_\text{FD}$). Since the valley-selective $|M_\sigma(\mathbf{k})|^2$ is maximum at the Dirac points, we restrict ourselves to photons with energies $E_\text{{ph}}=E_\text{{gap}}=2\Delta$. Thus, the intensity of optical absorption for a circularly polarized light is calculated from

\begin{multline}\label{eq_I}
I_\sigma (E_{\text{F}}) \propto \int_{BZ} |M_\sigma(\textbf{k})|^2 \delta(E_c(\textbf{k})-E_v(\textbf{k})-E_\text{{ph}}) \\ \times f_\text{FD}(E_v(\textbf{k}))\left(1-f_\text{FD}(E_c(\textbf{k}))\right) d{\textbf{k}}.
\end{multline}

Here, $E_\text{c}$ and $E_\text{v}$ are the energy of the conduction band and valence band, respectively. In this way, the circular dichroism at a given $E_{\text{F}}$ is expressed by the angle $\theta$

\begin{equation}\label{CD}
\tan \theta (E_{\text{F}})=\frac{I_+(E_{\text{F}})-I_-(E_{\text{F}})}{I_+(E_{\text{F}})+I_-(E_{\text{F}})},
\end{equation}
where $I_+$ and $I_-$ are the intensity of absorption for LCP and RCP, respectively. Although the optical matrix elements always present VP, this may not be the case for the optical absorption intensity. Therefore, we define the VP measure of the absorbed light as the difference of intensities between K ($I_\sigma^K$) and K$^\prime$ ($I_\sigma^{K^\prime}$) \cite{comment}

\begin{equation}\label{tanV}
\tan V_\sigma (E_{\text{F}})=\frac{I_\sigma^K(E_{\text{F}})-I_\sigma^{K^\prime}(E_{\text{F}})}{I_\sigma^K(E_{\text{F}})+I_\sigma^{K^\prime}(E_{\text{F}})}.
\end{equation}

In this case, when $\tan V_\sigma$ is $+1 (-1)$, light with circular polarization $\sigma$ is absorbed at the K (K$^\prime$) valley. We solve Eqs. (\ref{CD}) and (\ref{tanV}) for different values of $\Delta$, $t_2$ and $\phi$ at three different Fermi levels, identified by dashed lines in Fig. \ref{fig:fig1}(a) and Fig. \ref{fig:fig1}(b). When the Fermi energy lies at $E_\text{F}^\text{b}$ only electrons at the K valley can be excited with a photon energy of the band gap since the valence band maximum at K$^\prime$ is unoccupied. In this case, according to Fig. \ref{fig:fig1}(c), only RCP will be absorbed and therefore $\tan V_+(E_\text{F}^\text{b})=0$, $\tan V_-(E_\text{F}^\text{b})=+1$ and $\tan \theta(E_\text{F}^\text{b})=-1$. This is in sharp contrast with other hexagonal 2D materials or the Haldane model where VP and CD never occur simultaneously \cite{Ghalamkari2018}. Moving the Fermi energy to the conduction band ($E_\text{F}^\text{t}$) will restrict the light absorption to the K$^\prime$ valley since the conduction band at the K point is occupied. Hence, since the optical matrix elements at K$^\prime$ are opposite from those at the K valley, we obtain $\tan V_+(E_\text{F}^\text{t})=-1$, $\tan V_-(E_\text{F}^\text{t})=0$ and $\tan \theta(E_\text{F}^\text{t})=1$. Thus, changing the Fermi energy reverses both the CD and the VP. These two situations are the same for the semiconductor and the semimetallic phases, but differ in the case $E_\text{F} = E_\text{F}^\text{m}$, that is, in the midgap for the indirect semiconductor and crossing an electron and hole pocket for the semimetal regime. In this circumstance, the indirect semiconductor behaves like TMDs, where both valleys absorb opposite light polarization with vanishing CD. On the other hand, in the semimetallic band structure, no light is absorbed because the conduction band (valence band) is occupied (empty) at the K (K$^\prime$) point. 

\begin{figure}[t]
\begin{center}
\includegraphics[width=1 \columnwidth]{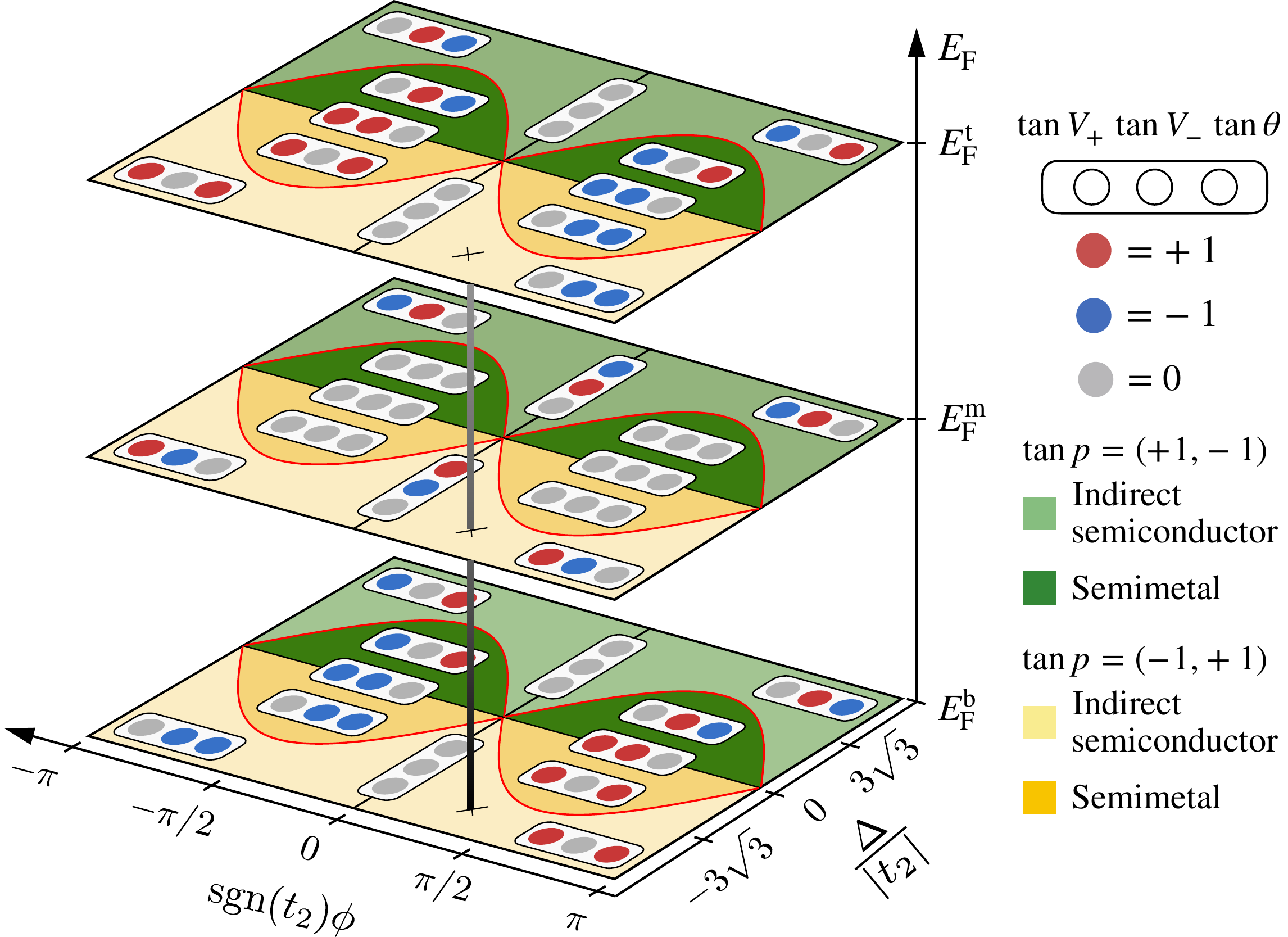}
\caption{Phase diagram of the modified Haldane model showing the rich combination of VP, CD and PP (see text for details).}
\end{center}
\label{fig:fig2}
\end{figure}

By studying all possible combinations of the modified Haldane parameters, we obtain a complex phase diagram showing a large variety of CD and VP possibilities, like a traffic signal, and different dependence with the Fermi level. This is depicted in Fig. 2, where $\tan V_\sigma(E_\text{F})$ and $\tan \theta(E_\text{F})$ are shown for the three selected Fermi energies (three planes in Fig. 2) and for all possible choices of $\Delta$, $t_2$ and $\phi$. To give a visual illustration of the results, we draw three colored circles for each region of the phase diagram. From left to right, the circles represent $\tan V_+$, $\tan V_-$ and $\tan \theta$ with values $+1,-1$ and $0$ for red, blue and grey colors, respectively. The plotted red line indicates the phase boundary from the semiconductor to semimetallic transition as is shown by the background colors, which in addition also display the pseudospin polarization. 

We observe that VP and CD occurs simultaneously at any point for $E_\text{F}^\text{b}$ and $E_\text{F}^\text{t}$ except at $\phi = 0$. In these regions, there is no change when going from a semiconductor to a semimetal, but the combination of VP and CD do vary when $\Delta$ or $\text{sgn}(t_2)\phi$ changes sign. Moreover, one can identify that the pattern in $E_\text{F}^\text{b}$ is the same as in $E_\text{F}^\text{t}$ but inverted with respect of $\text{sgn}(t_2)\phi$. The situation is different for the case of $E_\text{F}^\text{m}$. Firstly, now the semiconducting and semimetallic phases do not have the same optical properties. As mentioned above, no absorption occurs in the semimetallic case, leading to $\tan V_+ = \tan V_- = \tan \theta = 0$. However, the semiconducting band structure allows optical absorption, but since each valley absorbs opposite light polarization, CD vanishes. The valley polarization swaps when changing the sign of $\Delta$ due to PP inversion, as pointed out previously, but it remains the same when changing $\text{sgn}(t_2)\phi$.

It is also interesting to look at the transition points. These are the lines $\text{sgn}(t_2)\phi = 0$ and $\Delta/|t_2| = 0$. The former case is the situation of a 2D material with broken inversion symmetry, as in TMDs. In this case, since the Dirac cones are not shifted, the valence band maximum is empty at $E_\text{F}^\text{b}$ and the conduction band minimum is occupied for $E_\text{F}^\text{b}$. Thus, no absorption will occur for $E_\text{F}^\text{b}$ and $E_\text{F}^\text{t}$ even though there is both VP and CD for finite values of $\phi$. For $\Delta/|t_2| = 0$, the Dirac cones are shifted in energy but the system is always (semi)metallic with no energy gap. Nevertheless, we can promote optical absorption if we change the photon energy. If we take as an example $\text{sgn}(t_2)\phi > 0$, the K point is lowered down in energy. If we now place the Fermi level there (which we label as the $E_\text{F}^\text{b}$ case) such that no states near the K$^\prime$ are occupied and use small but nonzero photon energy, only electrons at K will absorb light. Since the PP is zero, $I_+(E_{\text{F}}) = I_-(E_{\text{F}})$ which leads to the peculiar case $\tan V_+ = \tan V_- = 1$ and $\tan \theta = 0$. Finally, for the special case of $\text{sgn}(t_2)\phi = \Delta/|t_2| = 0$, corresponding to graphene, we find that no matter which $E_{\text{F}}$ and $E_\text{ph}$ we choose, there is not PP, VP or CD.

\begin{table}[t]%
\centering
\label{tab_Haldane_hop}
\begin{tabular}{cccc} \hline
PP & VP  & CD & Model/Material \\ \hline
$\times$ & $\times$ & $\times$ & Graphene, ${\rm H \;  \& \ mH^{\rm a}}$ \\
$\times$ & $\times$ & $\bigcirc$ & - \\			
$\times$ & $\bigcirc$ & $\times$ & ${\rm H^{\rm b} , mH^{\rm c}}$ \\
$\times$ & $\bigcirc$ & $\bigcirc$ & - \\
$\bigcirc$ & $\times$ & $\times$ & ${\rm mH^{\rm d}}$ \\
$\bigcirc$ & $\times$ & $\bigcirc$ & ${\rm H^{\rm e}}$ \\
$\bigcirc$ & $\bigcirc$ & $\times$ & hBN, TMD, ${\rm H^{\rm f}, mH^{\rm g}}$\\
$\bigcirc$ & $\bigcirc$ & $\bigcirc$ & ${\rm mH^{\rm h}}$ \\ \hline
\end{tabular}
\caption{Combination of presence ($\bigcirc$) and absence ($\times$) of pseudospin, valley polarization and circular dichroism and the materials/models where such combination occurs. The Haldane/modified Haldane model is abbreviated as H/mH. Parameters are ${\ }^{\rm a}: \Delta = \phi = 0, E_\text{F}^\text{m} (\text{semimetal}, \Delta=0)$; ${\ }^{\rm b}:\Delta/t_2 = 3\sqrt{3}$; ${\ }^{\rm c}:E_\text{F}^\text{b,t} (\Delta=0)$; ${\ }^{\rm d}:E_\text{F}^\text{m} (\text{semimetal}, \Delta\neq 0)$, $E_\text{F}^\text{b,t} (\phi=0)$  ; ${\ }^{\rm e}$: topological insulator; ${\ }^{\rm f}$:trivial insulator; ${\ }^{\rm g} :E_\text{F}^\text{m}$ (indirect semiconductor); ${\ }^{\rm h}: E_\text{F}^\text{b,t} (\Delta \neq 0, \phi \neq 0)$.}
\end{table}

We summarize all possible combinations of presence and absence of PP, VP, and CD in Table 1. As novel phases, ${\rm mH}^{\rm d}$ gives access to an exclusive PP ($\bigcirc$) in absence of VP and CD ($\times$). Differently, ${\rm mH}^{\rm h}$ describes the situation with nonzero $\Delta$ and $\phi$ values and for Fermi energies $E_\text{F}^\text{b}$ and $E_\text{F}^\text{t}$, which permits all polarizations and CD to be active. We note that is not possible to obtain CD in absence of PP, since whenever PP vanishes, $I_+(E_{\text{F}}) = I_-(E_{\text{F}})$ for both valleys, which implies that we cannot achieve CD. 

\begin{figure}[t]
\includegraphics[width=1 \columnwidth]{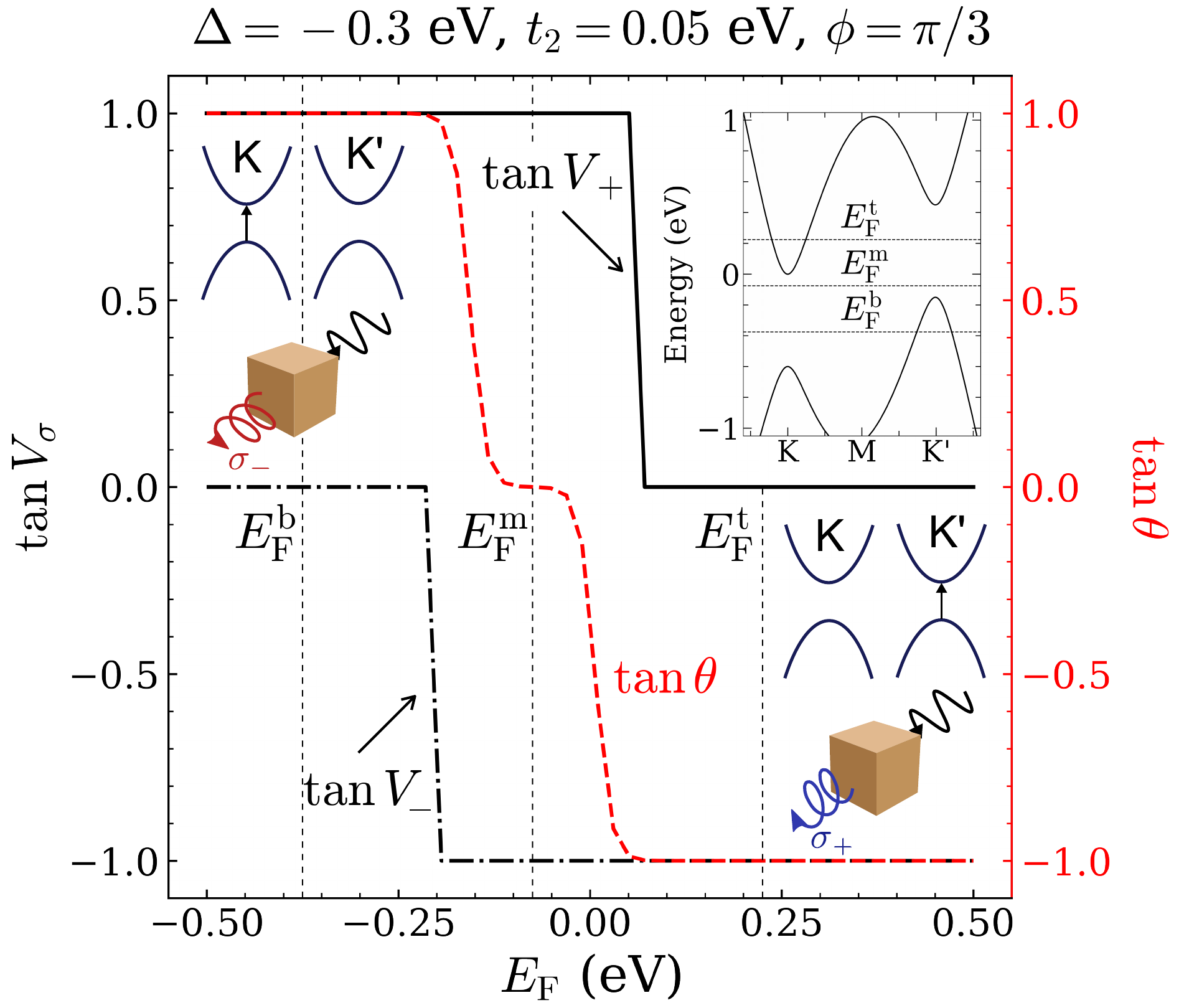}
\caption{VP and CD as a function of the Fermi energy for the case corresponding to the cross marked in Fig. 2. The insets illustrate the band structure for these parameters in addition to the proposed mixed device functionality: a light polarizer and valley selective excitation.}
\label{fig:fig3}
\end{figure}

Additionally, the ${\rm mH}^{\rm h}$ presents a particularly unique feature which is the tunability of both VP and CD simultaneously. Fig. \ref{fig:fig3} illustrates such property for varying the Fermi energy for a selected case in the phase diagram (marked with a cross in Fig. 2 and connected vertically with a grey line). This variation of the Fermi energy could be  performed in an experiment by applying an external gate voltage. The resulting absorption profile of \textit{perfect} PP, VP and CD suggests interesting technological applications. Indeed, by combining perfect VP and CD, one could envision a device that can be useful for valleytonics and simultaneously serve as a tunable light polarizer (as illustrated in Fig. \ref{fig:fig3}, insets). At low Fermi energies, and using linear polarized light, the material would absorb the LCP component at the K valley and consequently transmitting RCP light. These excited electrons could be driven into a circuit as a source of valley current and at the same time the RCP light could be used for any application requiring circularly polarized light. By increasing the Fermi energy, we could reverse the device functionality and thus create valley current at the K$^\prime$ valley while transmitting LCP light. The possibilities could go even further: by having regions of the device with $E_\text{F} = E_\text{F}^\text{b}$ and other parts with $E_\text{F} = E_\text{F}^\text{t}$, we could create valley currents with LCP light and use the transmitted RCP component to generate opposite polarized valley currents in the other device region.

In summary, we have presented fascinating optical properties of the modified Haldane model, which displays a rich phase diagram showing tunable pseudospin and valley polarization and circular dichroism. The realization of both perfect valley polarization and circular dichroism enable the concept of integrating photonics and valleytronics in a single material or device. Moreover, this suggests an optical readout for discovering materials in which antichiral edge states could form \cite{Colomes2018}.

\begin{acknowledgments}
M. Vila acknowledges the Graphene Flagship grant and the Department of Physics of Tohoku University for its hospitality. The research leading to these results has received funding from $"$La Caixa$"$ Foundation by supporting M. Vila. S. Roche was supported by the European Union Horizon 2020 research and innovation programme under grant agreement No. 696656 (Graphene Flagship). ICN2 is funded by the CERCA Programme / Generalitat de Catalunya, and is supported by the Severo Ochoa program from Spanish MINECO (Grant No. SEV-2017-0706). R. Saito acknowledges JSPS Kakenhi (Grant No. JP18H01810). N.T. Hung acknowledges JSPS Kakenhi (Grant No. JP18J10151).
\end{acknowledgments}


%

\end{document}